\documentclass[graybox]{svmult}

\usepackage{mathptmx}       
\usepackage{helvet}         
\usepackage{courier}        
\usepackage{type1cm}        

\usepackage{makeidx}         
\usepackage{graphicx}        
\usepackage{multicol}        
\usepackage[bottom]{footmisc}

\usepackage{url}

\makeindex             

\begin{document}

\newcommand{\fig}[2]{
 \begin{figure}[t]
    \centering
    \includegraphics[width=1.0\linewidth,clip]{#1.png}
    \caption{#2}
    \label{fig:#1}
  \end{figure}}

\title*{Modeling Internet-Scale Policies for Cleaning up Malware}

\author{Steven Hofmeyr \and Tyler Moore \and Stephanie Forrest \and Benjamin
  Edwards \and George Stelle}
\authorrunning{Steven Hofmeyr et al.}
\institute{  
  Steven Hofmeyr \at Lawrence Berkeley National Laboratory, \email{shofmeyr@lbl.gov} 
  \and
  Tyler Moore \at Harvard University, \email{tmoore@seas.harvard.edu} 
  \and 
  Stephanie Forrest \at University of New Mexico, \email{forrest@cs.unm.edu}
  \and
  Benjamin Edwards \at University of New Mexico, \email{bedwards@cs.unm.edu}
  \and
  George Stelle \at University of New Mexico, \email{stelleg@cs.unm.edu}}

\maketitle

\abstract*{
  An emerging consensus among policy makers is that interventions undertaken by
  Internet Service Providers are the best way to counter the rising incidence of
  malware.  However, assessing the suitability of countermeasures at this scale
  is hard. In this paper, we use an agent-based model, called ASIM, to
  investigate the impact of policy interventions at the Autonomous System level
  of the Internet. For instance, we find that coordinated intervention by the
  0.2\%-biggest ASes is more effective than uncoordinated efforts adopted by
  30\% of all ASes.  Furthermore, countermeasures that block malicious transit
  traffic appear more effective than ones that block outgoing traffic.  The
  model allows us to quantify and compare positive externalities created by
  different countermeasures.  Our results give an initial indication of the
  types and levels of intervention that are most cost-effective at large scale.
}

\section{Introduction}
\label{sec:introduction}

Many Internet-connected computers are infected with malicious software, or {\em
  malware}.  Malware can harm the infected computer user directly, for example,
by installing a keystroke logger to collect confidential information
surreptitiously. It can also place the machine into a botnet consisting of
thousands or even millions of computers that carry out attacks of the operator's
choosing, such as sending email spam or launching denial-of-service attacks.
Infected machines can also become vectors for further malware spread, as in the
case of Conficker, which initiates attacks from infected machines to recruit new
computers to the botnet~\cite{SRIConficker}.

In economic terms, malware imposes negative externalities by harming innocent
third parties~\cite{AndersonMoore06}.  Negative externalities are a form of
market failure, which suggests that there will be an oversupply of the resource
(in this case, malware) in equilibrium.  Policy makers are interested
in correcting this market failure to reduce the social cost of malware.
Although many stakeholders could potentially help control the spread of
malware, the emerging consensus is that Internet Service Providers (ISPs) are
best positioned to intervene~\cite{MooreJEP09,AndersonWEIS08,VanEetenWEIS10}.

It is less clear, however, what kind of intervention is most appropriate. The
possibilities range from simply notifying infected customers to actively
quarantining them until the malware has been demonstrably removed.  It is
difficult to gauge the impact of policies and ISP-level interventions until they
have been tried, and it is expensive (both financially and in terms of political
capital) to adopt industry-wide policies.  Consequently, it is important to get
it right the first time.

One way to address this issue is through modeling.  In this paper we model
potential intervention strategies for controlling malware and compare their
likely impact.  We use an agent-based model called ASIM~\cite{HolmeEtAl08a},
which represents the Internet at the autonomous system (AS) level, the level at
which policy interventions are being actively considered.  ASIM incorporates
traffic, which is key to understanding the spread of malware, geography, which
is key to investigating country-level effects, and economics, which is is key to
understanding the cost and benefits of interventions.

Through a series of experiments we study several questions, reporting some
findings that are unsurprising and others that are counterintuitive. For
example, our experiments show, as we would expect, that a few of the largest
ISPs acting in concert are more effective than a randomly chosen subset of all
ASes intervening unilaterally. However, the numbers involved are more
surprising: Intervention by the top $0.2$\% of ASes is more effective than
intervention by 30\% of ASes chosen at random. Our results also suggest that
when only the largest ASes intervene, it is better to simply filter out
malicious traffic (especially transit traffic) than to attempt to remediate
end-user infections. We also explore briefly the impact of interventions on the
growth of the network, and demonstrate that policies that are beneficial in the
short term could be harmful in the long-term.  For example, the collateral
damage caused by blacklisting malicious traffic sources promotes those ASes that
profit from receiving more malicious traffic.

The remainder of the paper is structured as follows. We review in greater detail
the policy interventions currently under consideration worldwide in
Section~\ref{sec:policy}.  In Section~\ref{sec:model}, we explain how ASIM works
and how the cybersecurity interventions are implemented.  In
Section~\ref{sec:validation} we describe how we empirically validated ASIM, and
Section~\ref{sec:exp} reports experimental results.  We discuss related work in
Section~\ref{sec:related} and the findings and limitations in
Section~\ref{sec:discussion}.  Finally, we conclude in
Section~\ref{sec:conclusion}.

\section{Policy Interventions}
\label{sec:policy}

There are several reasons why ISPs are a promising point of intervention.
First, ISPs are the gatekeeper to the Internet for many computers and thus in a
unique position to inspect traffic to and from their customers.  Infections are
often detected remotely by scanning for outgoing connections to known
command-and-control servers used by botnet operators~\cite{MAAWG09}.  In this
scenario, only the ISP can link an IP address to customer details, a crucial
step if customers are to be notified and assisted.

A second reason is that ample opportunity exists for reducing the prevalence of
malware by enlisting the help of ISPs.  Using several years' worth of data on
computers sending spam (a natural proxy for botnet activity), van Eeten et
al.~\cite{VanEetenWEIS10} found that most compromised computers were customers
of legitimate ISPs, and that infection rates vary dramatically across ISPs and
countries.  Their evidence suggests that differences in security
countermeasures, not merely target selection by attackers, can affect infection
rates at ISPs.

However, incentives for ISPs to implement security countermeasures are weak.  As
mentioned above, much of the harm caused by malware is externalized, but the
cost of intervention would fall largely on the ISP.  Although the infected host
is often unharmed by malware, the ISP is definitely not directly harmed.
However, the cost of notification and cleanup can be substantial. According to
an OECD study, one medium-sized ISP reported that it spent 1--2 \% of its total
revenue handling security-related support calls~\cite{VanEeten08}.  Thus, there
is a strong disincentive for ISPs to notify infected customers and also pay for
any resulting support calls.

Despite weak incentives, ISPs in many countries have begun exploring a variety
of remedial interventions, either with government cooperation or to preempt the
imposition of more burdensome regulatory requirements. Interventions by ISPs
usually do not include the detection of malware, only remediation once malware
is detected. For notifications of misbehaving or compromised customers, ISPs
rely on third parties, such as the operators of email blacklists, botnet
trackers, other ISPs and security companies,

Once a threat is identified, most ISPs choose to do nothing, waiting until the
abuse team has time to act or for additional warnings about the customer to
accrue.  However, some ISPs have begun to notify customers.  In the US, Comcast
automatically notifies customers of infections with a browser pop-up that links
to instructions for removing the malware~\cite{Comcast}.  The customers are
responsible for completing the clean-up process, and it is inevitable that not
all malware will be removed successfully even after notification.  As a further
step, Comcast has partnered with Symantec to offer remediation by a skilled
technician for \$100.  A similar approach is being rolled out by Australian
ISPs~\cite{AussieISP}.

A more aggressive step is to place infected computers into ``quarantine.''  Once
in quarantine, users are required to download and install anti-virus software
and malware removal tools.  They leave the quarantine only after the security
software is installed and the computer passes a network-based scan for malware.
Quarantine is considerably more expensive than the notification-only approaches,
and the the ISPs that use them do so only for a minority of affected customers.
Recently, the Dutch ISPs announced a signed agreement to notify and quarantine
affected customers~\cite{DutchISP}.

Both ISPs and policy makers have realized that tackling widespread infection can
be made more effective if ISPs coordinate their interventions.  In both the
Dutch and Australian case, many ISPs have joined together in common action,
prodded by their governments. This collective action is designed in part to
allay the fear that customers might switch providers rather than fix the
underlying problem.

Some countries are weighing more active intervention.  If the cost of customer
support is really the greatest impediment to ISP action, then the German
government's decision to establish and subsidize a nationwide call center could
really help~\cite{GermanyCall}.  Under this plan, ISPs will identify infected
customers and pass along the information to the call center.  Clayton describes
a proposal under consideration by Luxembourg to subsidize the cost of voluntary
cleanup whenever a customer has been notified of infection~\cite{ClaytonWEIS10}.
Instead of such ``carrot''-based incentives, ``sticks'' could also be tried.
Anderson et al. recommended that the European Commission introduce fixed
penalties for ISPs that do not expeditiously comply with notifications of
compromised machines present on their networks~\cite{AndersonWEIS08}.

Finally, policy makers could coordinate their defenses by aggregating
notifications of infection.  A survey of Dutch ISPs revealed that they notify or
quarantine only about 10\% of infected customers~\cite{VanEetenDutch10} even
though they claim to notify all customers known to be infected.  This occurs
because their individual lists of infections are incomplete.  Data
incompleteness is a widespread problem in information
security~\cite{MCEcrime08}, as firms often jealously guard their incident
information as trade secrets.  To combat this trend, the Australian Internet
Security Initiative now aggregates data on compromised machines into a single
feed and passes it along to Australian ISPs~\cite{AussieISP}.

\section{Model Description}
\label{sec:model}

ASIM~\cite{HolmeEtAl08a} is an agent-based model of Internet growth at the
Autonomous System (AS) level. ASes roughly correspond to ISPs. While there are
differences between ASes and ISPs (e.g., a single ISP can use several AS
numbers), more extensive and reliable data is available describing ASes than
ISPs. This eases empirical validation and explains why most of the literature
has studied Internet topology at the AS level. We summarize the important
features of ASIM here, highlighting differences between the original
implementation and the version used in this paper.

ASIM is based on highly simplified implementations of four key features of ASes:
network structure, traffic flow, geography, and economics. These features are
sufficient to enable ASIM to generate networks with topologies, dynamics, and
spatial distributions similar to those of the Internet. There are conceptual
similarities between ASIM and some earlier Internet models such as
HOT~\cite{peer:chang,chang:superhot}, although many of the details are
different.  For example, ASIM adds explicit economic considerations and accounts
directly for population density. 

ASIM attempts to reproduce large-scale features of the AS level of the Internet
by modeling localized and well-understood network interactions. Instead of
simply reproducing a macroscopic pattern using statistical fitting or
phenomenological models, ASIM specifies a set of primitive components (the
agents) and interaction rules that mimic the architecture of the real
system. The model is run as a simulation, and macroscopic behaviors (e.g.,
degree distribution) are observed and compared to real-world data. The objective
is to provide a parsimonious explanation of how a system works by hypothesizing
a small set of simple but relevant mechanisms.

In ASIM each AS is an economic agent, which manages traffic over a
geographically extended network (referred to as a {\em sub-network} to
distinguish it from the network of ASes) and profits from the traffic that flows
through its network. We assume a network user population distributed over a
two-dimensional grid of locations. Traffic is generated between source and
destination with a probability that is a function of the population profile. The
model is initialized with one agent that spans one grid location. At each time
step a new agent is added to a single location. As time progresses, each agent
may extend its sub-network to other locations, so that the sub-networks reach a
larger fraction of the population. This creates more traffic, which generates
profit, which is then reinvested into further network expansion. In addition,
agents link to each other, potentially routing traffic between sub-networks other
than their own. A necessary, but not sufficient, condition for two agents to be
connected is that they overlap in at least one location. Through positive
feedback, the network grows until it covers the entire population.

For this paper, we have reimplemented ASIM in order to make it run efficiently
in parallel.\footnote{Code available at \url{http://ftg.lbl.gov/projects/asim}.}
In the process, we have simplified the model, without reducing the accuracy with
which the model simulates AS-like networks. The major changes are described
below.

\subsection{Simplifying the Original ASIM}

In the original model described in Holme et al.~\cite{HolmeEtAl08a}, a variable
number of agents could be added every time step, sufficient to maintain the
correct average degree. In the new model, we simply add one agent per iteration,
regardless. This follows realistic observed growth curves where the number of
new agents grows at an almost perfectly linear rate. In our analysis of the real
world data, we find that about $5.5$ new ASes are added per day, so in our
simulation, one time step is the equivalent of approximately $4.4$ hours. Each
new agent is added to a single, already occupied location\footnote{Except for
  the very first agent, of course.}, chosen at random (weighted according to
population).

Instead of a packet-switched model, we use the gravity
model~\cite{gravityModel}. For the gravity model, the traffic flow $T$ between a
pair of agents A and B is
$$T(A,B) = \frac{pop(A)pop(B)}{d(A,B)^2}$$
where, $pop(A)$ is the population served by A, $pop(X)$ is the population served
by B, and $d(A,B)$ is the shortest path distance on the AS graph from A to
B. Once we have determined the flow between A and B, we propagate it across the
graph on the shortest path and every agent along that path gets its count of
traffic increased accordingly. If there are multiple shortest paths, we randomly
choose one. This traffic flow computation is performed for every pair of agents.

The traffic model is run every 16 time steps, corresponding to every three days
of simulation time. Computing paths and carrying out traffic flow is expensive
and most paths do not change significantly in the short term. We find
experimentally that running the traffic model every 16 time steps provides a
good balance between computational overhead and maintaining accuracy. Note that
there is no notion of capacity, as there was in the original model.

There are two major differences in the modeling of geography. First, we
disregard geographic distance, i.e. the cost of expanding to a new location is
constant, regardless of where an agent expands to. By contrast, in the
original model, the greater the distance from an agent's existing locations to a
new location, the higher the cost of expansion. Second, in the new ASIM, an
agent expands to a randomly chosen location, weighted by populace, regardless of
how many other agents exist at that location. This differs from the
original model, where the location chosen was the one with the highest
shared\footnote{The population of the location, divided by the number of agents
 with presence at that location.} population within reach.

The mechanism for earning revenue in the new implementation is very similar to the
original model. In the original model, an agent earns money for every packet it
transits. In the new ASIM, we do not have a packet-switched model, and so an
agent simply earns money every iteration proportional to the volume of traffic
that it transits in either direction.

It does not cost an agent to link, unlike in the original model. There are two
circumstances in which new links are added. First, when a new agent is placed
at a location, it is linked to an agent that is chosen uniformly at random from
those already at that location. This ensures the graph remains
connected. Second, as in the original model, a number of links is added on every
iteration, sufficient to maintain the desired average degree. In this case, when
a link is added, the source is chosen uniformly at random from all agents, and
the destination is chosen by first choosing an occupied location (weighted
according to population), and then selecting uniformly at random one of the
agents at that location. If the source does not exist at that location, it 
expands to that location. This ensures that agents can only link if they share a
location, as in the original model.

\subsection{Adding Cybersecurity to ASIM}
\label{sub_sec:AddingCybersecurityToASIM}

We use ASIM to compare the effectiveness of different policy interventions
that counter the proliferation of malware infections. For simplicity, we
assume that every AS can implement interventions, i.e. we do not focus on ISPs
alone. We define insecurity by assigning a \emph{wickedness rate} to each AS:
the fraction of machines that are infected with malware. Depending on its size,
each AS has a corresponding {\it wickedness level}: the absolute number of
infected machines.  Sometimes we will simply refer to wickedness as an
abbreviation of wickedness level. We define the wickedness rate $w_i$ for each
AS $i$ according to the exponential distribution:
\[w_i = \mathrm{min}(-\overline{w} \ln(1 - r_i)),0.5)\]
where $r_i$ is a value selected uniformly at random from the interval
$[0,1]$, and $\overline{w}$ is the average wickedness. In
Section~\ref{sec:validation} we explain why this distribution is a
reasonable match to observed empirical measurements of wickedness.

In ASIM, the \emph{wicked traffic} that flows from a source AS A to a
destination AS B is directly proportional to the wickedness level at A. We
define the {\it wicked traffic rate} at B as the fraction of all traffic
destined for end users at B that is wicked. Hence we do not count transit
traffic when measuring wickedness, although wicked traffic is passed through the
network. We are only interested in the impact of wicked traffic on end users,
and so are only concerned with the volume of traffic that reaches the
destination.

We model five types of interventions that can be undertaken by each AS:
\begin{enumerate}
\item {\bf Do nothing}: This is the baseline where the AS makes no active
 intervention.
\item {\bf Reduce egress wickedness}: This captures a range of AS interventions
 that remediate customer infections. The percentage reduction of wicked egress
 traffic depends on the aggressiveness of the intervention---automated
 notifications are less successful than quarantine, etc.
\item {\bf Reduce ingress wickedness}: An AS can deploy filters that drop some
 portion of incoming wicked traffic. The proportion dropped depends on the
 effectiveness of wicked traffic detection, the capacity of filtering on the
 routers, and other factors. Ingress filtering can be applied to both end-user
 traffic and transit traffic.
\item {\bf Reduce egress and ingress wickedness}: An AS can deploy methods 2 and 3
 simultaneously.
\item {\bf Blacklist wicked traffic sources}: An AS can drop all traffic originating
 from known wicked sources, typically dropping all traffic that comes from
 another AS that is known to have high infection rates. Hence there is
 collateral damage because legitimate as well as wicked traffic is dropped. We
 model this by having an AS drop all traffic (both wicked and legitimate) from
 other ASes with sufficiently high wickedness rates. We also model the notion
 of an AS being \emph{too big to block}, i.e. an AS will only blacklist smaller
 ASes because blacklisting large ASes is expected to result in an excessive
 loss of legitimate traffic.
\end{enumerate}

Another intervention under consideration by policy makers is increased
\emph{data sharing}, where an AS learns about infections from an amalgamation of
sources. We do not treat data sharing as a separate intervention in the model;
rather, we can observe the effect of increased data sharing by increasing the
effectiveness of ingress and egress interventions.

Separately, we model which ASes choose to intervene as follows:
\begin{enumerate}
\item {\bf Unilateral}: Some ASes choose to intervene unilaterally, and there is
 no coordination between ASes or regulatory pressure on a particular subset of
 ASes to intervene. We implement this by randomly selecting a subset of ASes
 to adopt intervention strategies.
\item {\bf Large ASes act in concert}: A selection of large ASes together adopt
 one of the AS-level interventions. There are several variations on this:
\begin{enumerate}
\item {\it Global coordination}: All the largest ASes adopt one of the AS-level
 interventions.
\item {\it Country-specific coordination}: All of the largest ASes in one
  country adopt one of the AS-level interventions. We implement this in the
  model by randomly selecting a fraction of the largest ASes to apply
  interventions.
\item {\it Small AS inclusion}: Smaller ASes also adopt the interventions.
\end{enumerate}
\end{enumerate}

\section{Validating the Model}
\label{sec:validation}

The original ASIM~\cite{HolmeEtAl08a} was validated on real world data and shown
to be a close match on a number of metrics. That work dates from 2006, so we
have collected more recent data to perform more extensive validation of the new
ASIM. First, we gathered data on the real topology of the AS graph using the
standard method of inferring links from BGP dumps, which we collected from the
RouteViews\footnote{\url{www.routeviews.org}} and
RIPE\footnote{\url{www.ripe.net}} databases. These data were used to validate
ASIM on 12 different graph-based metrics; the results are too extensive to
include in this paper.\footnote{Data and tools available at
  \url{http://ftg.lbl.gov/projects/asim}.}

Second, we gathered data on the distributions of locations among ASes in the
real world by matching geoip information from
MaxMind\footnote{\url{www.maxmind.com}} with the IP prefixes of ASes collected
from the BGP dumps. We used this data to confirm that the characteristics of the
geographical distribution of agents in ASIM correspond closely with the real
Internet. We also used MaxMind to gather population data for cities matched to
locations inferred from the geoip data. We could thus confirm that the
characteristics of the population distribution in ASIM closely follow that in
the real world.

Obtaining data to validate the cybersecurity extensions to ASIM is a more
challenging task. Reliable data are difficult to find for the most important
quantity: the distribution of wickedness rates over the ASes. Perhaps the best
data comes from a study by Van Eeten et al.~\cite{VanEetenDutch10} of botnet
activity at Dutch ISPs. The authors aggregate data on IP addresses observed
to be sending email spam, participating in the Conficker botnet, or appearing in
the logs of intrusion detection systems for suspected attack behavior. They
found that between 2\% and 7\% of the customers of the nine largest Dutch ISPs
were infected and exhibiting botnet activity.

Van Eeten et al. also collected similar data on global Internet activity,
finding that Dutch ISPs experience slightly lower than average rates, with the
worst-performing countries experiencing a rate several times higher than that of
of the Dutch ISPs. However, the authors do not report rates for other countries,
because some countries make more extensive use of DHCP than the Netherlands,
which could lead to overestimates. To incorporate the potential for higher
rates, for our experiments we selected an average wickedness rate
$\overline{w}=0.1$, slightly higher than the highest Dutch ISP value.

Although we can derive the average wickedness rate from the Dutch data, we are
also interested in how wickedness is distributed across ISPs. To that end, we
collected per ISP data from two sources of malicious activities. First, we
collected data from \url{maliciousnetworks.org}, where academic researchers have
constructed a system that tallies the level of malicious activity at each
AS~\cite{fire}. They aggregate reports of botnet, phishing and malware servers
observed at each AS. Second, we analyzed a single-day snapshot from the SANS
Internet Storm Center, which publishes a list of over 1 million IP addresses
exhibiting attack
behavior~\footnote{\url{http://isc.sans.edu/feeds/daily_sources}}. We then
determined the AS associated with each IP address in the SANS list and tallied
the total number of IP addresses observed at each AS to arrive at measures of
wickedness levels for the ASes. Note that in both of these cases, we can
determine only wickedness levels, not rates, because the number of customers
served by each AS is not publicized.

Figure~\ref{fig:Fig1} plots the complementary
cumulative distribution function (CCDF) of wickedness levels obtained from
\url{maliciousnetworks.org}, the Internet Storm Center, and ASIM. We can see
that our use of an exponential distribution for the wickedness levels in ASIM
results in a simulated CCDF that falls between the two empirical data sets. From
this, we conclude that the method used in ASIM for generating wickedness rates
for ASes is reasonable.

\fig{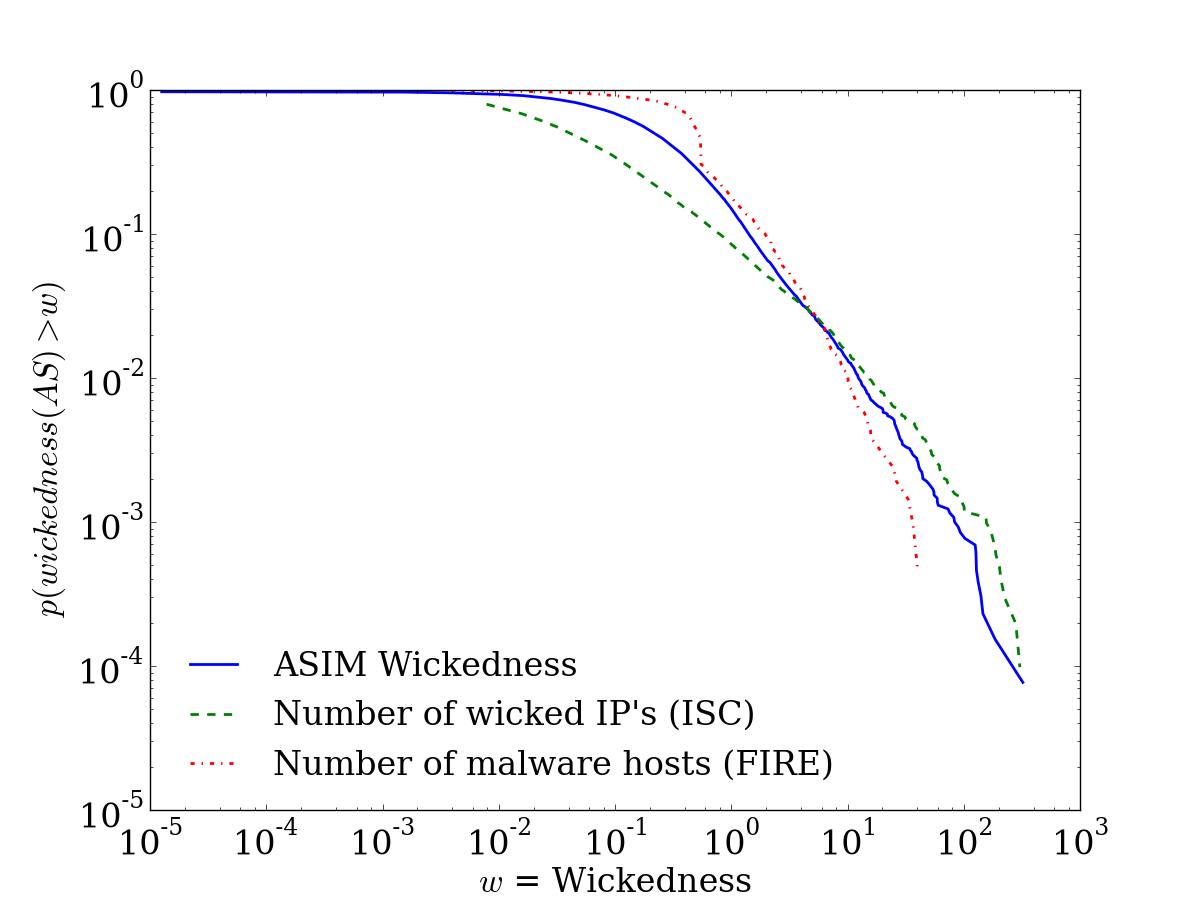}{The distribution of wickedness levels generated by ASIM and in two
  real world data sets. (Normalized)}

Even less data are available to evaluate the effectiveness of the different
policy interventions described in Section~\ref{sec:policy}. To our
knowledge, the only data on interventions comes from the same Dutch study
mentioned above~\cite{VanEetenDutch10}. The authors surveyed ISPs about how
often they notified or quarantined customers infected with malware, and then
compared this to their own measurements of wickedness levels. They found that
ISPs notified between 1\% and 50\% of infected customers, and that around
20-25\% of this number were also placed into quarantine. As a baseline, in ASIM
we assume that standard intervention reduces wicked traffic by 20\%, although in
Section~\ref{sec:exp}, we also explore the impact of varying the remediation
efficacy. We place the different intervention techniques on a continuum:
notification is less effective than quarantine, and both can be substantially
improved by sharing notifications.

\section{Experimental Results}
\label{sec:exp}

\fig{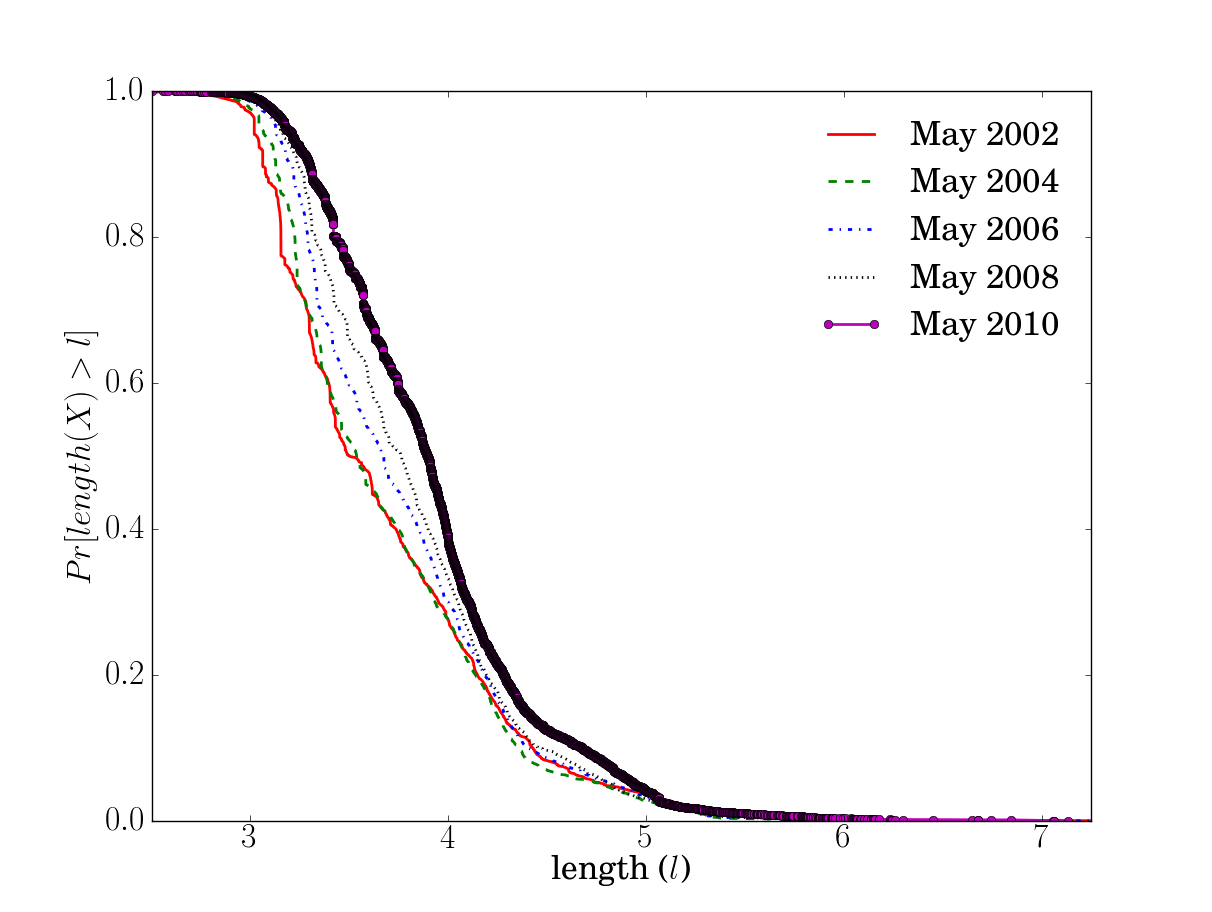}{The change over time of the complementary cumulative
 distribution (CCDF) for the average path length between every pair of ASes in
 the real Internet.}

We carried out a number of experiments to explore the impact of the various
cybersecurity interventions modeled in ASIM. First, in
Section~\ref{sub_sec:ImpactAtASingleInstant}, we investigate the simulation at a
single point in time, and second, in Section~\ref{sub_sec:ImpactOnNetworkGrowth}
we study the simulation as the network evolves. In both cases, we measure the
impact of an intervention as the percentage by which it reduces the wicked
traffic rate (as defined in Section~\ref{sub_sec:AddingCybersecurityToASIM})
compared to when no intervention is adopted. When interventions occur, they
filter out 20\% of wicked traffic, except for blacklisting, where all traffic
from a blacklisted AS is dropped, both legitimate and wicked. For all
experiments, we used the default parameter settings for ASIM
V0.3.\footnote{\texttt{av\_degree = 4.2}, \texttt{extent\_cost = 1.5},
  \texttt{base\_income = 5}, \texttt{pop\_distr\_exp = -1}, \texttt{wickedness =
    0.1}.}

\subsection{Impact at a Single Instant}
\label{sub_sec:ImpactAtASingleInstant}

For our study of the effect of interventions at a single point in time, we used
ASIM to grow a network of 10\,000 ASes, and used that network as the basis for
all experiments. For each intervention, we started with the same 10\,000 AS
network, set the parameters appropriately, and ran ASIM for a single time
step. The traffic component of ASIM always updates at the end of a run, so this
yields a single update of the traffic patterns, changed according to the
intervention, and always starting from the same state.

We used 10\,000 ASes, rather than the current approximately 34\,000 in the real
Internet,\footnote{As of May 2010.} to reduce the running time of the
simulation. This should have no substantive impact on the experimental results
because the key characteristics of the AS-level graph do not change
significantly as the network grows, either in our simulations or in reality. For
example, Figure~\ref{fig:Fig2} shows that the distribution of
average path lengths has remained roughly unchanged over the last decade,
even as the number of ASes has grown more than threefold.

\fig{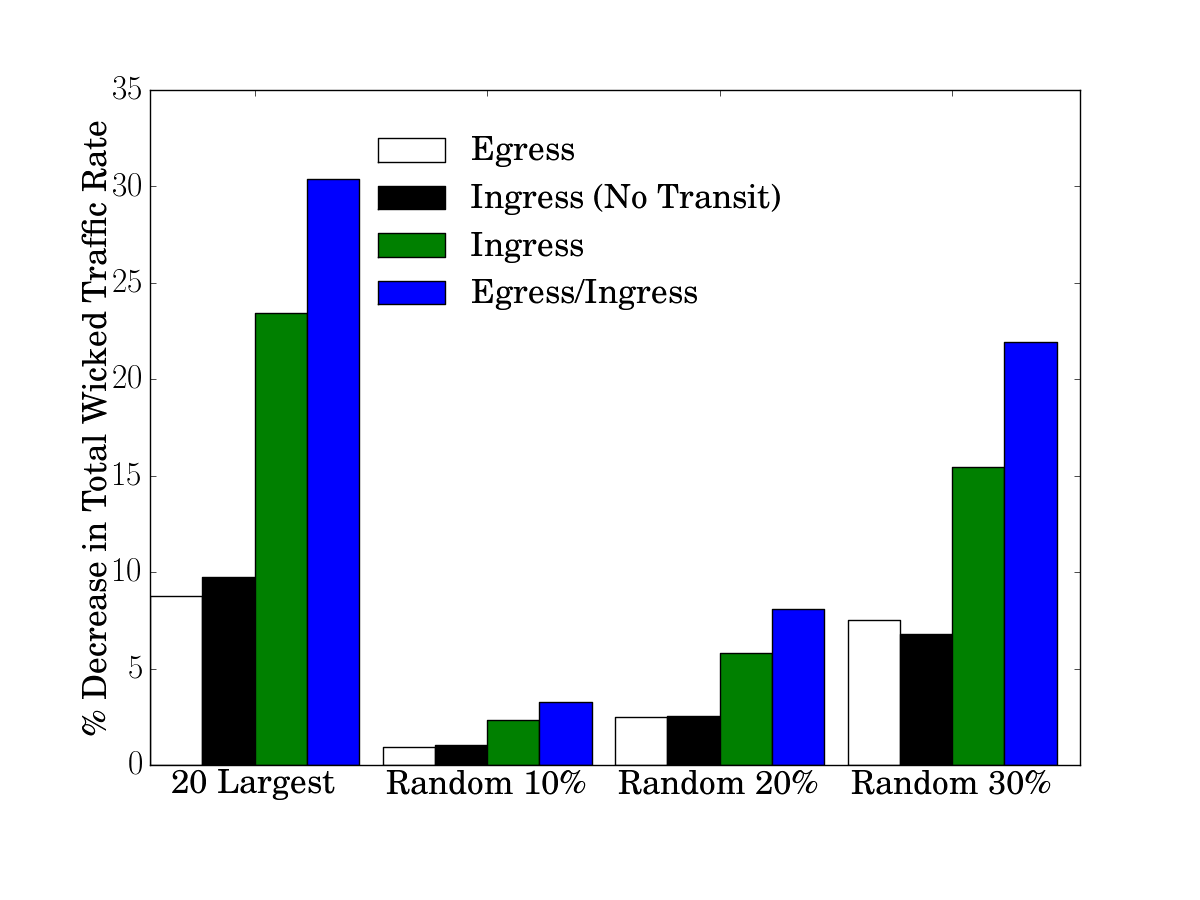}{The impact of interventions on wicked traffic rate. ``20 largest'' is
  the effect when the 20 largest ASes intervene; ``random $x$\%'' is the effect
  when $x$ percent of all ASes intervene.}

We first examine how applying interventions to different ASes can affect wicked
traffic levels. Figure ~\ref{fig:Fig3} shows how
wicked traffic decreases when only the 20 largest ASes (as measured by degree)
adopt interventions, as compared to a random selection of between 10-30\% of all
ASes. This illustrates the case where interventions are coordinated at the largest
ISPs to a hands-off approach where ISPs decide for themselves whether or not to
adopt countermeasures. The graph clearly demonstrates that targeting the largest
ASes is a superior strategy, given that targeting just the 20 largest ASes
($0.2$\% of the total) reduces traffic by more than applying interventions to
even 3\,000 randomly selected ASes.

It is not particularly surprising that targeting the largest ASes is the most
effective strategy, given the structure of the AS graph. In our simulations, the
largest ASes route up to six orders of magnitude more traffic than the
smallest. Nonetheless, the results reinforce the argument that remediation
policies can be more successful by focusing on a small group of the largest
ASes, unless a majority of all ASes can be persuaded to unilaterally respond.

What is more striking is the comparison between ingress and egress filtering.
Filtering ingress traffic destined for end users only (i.e. not filtering
transit traffic) is about as effective as filtering egress traffic (around 10\%
when the largest ASes intervene). Ingress filtering of both end-user and transit
traffic at the largest ASes, by contrast, reduces wicked traffic by a factor of
$2.7$ over egress alone. This is a more surprising finding, as it suggests that
filtering incoming wicked traffic is more effective than stopping outgoing
traffic. When ASes act unilaterally, the difference is not as large (a factor of
$1.8$) because the smaller ASes transit less traffic.

\fig{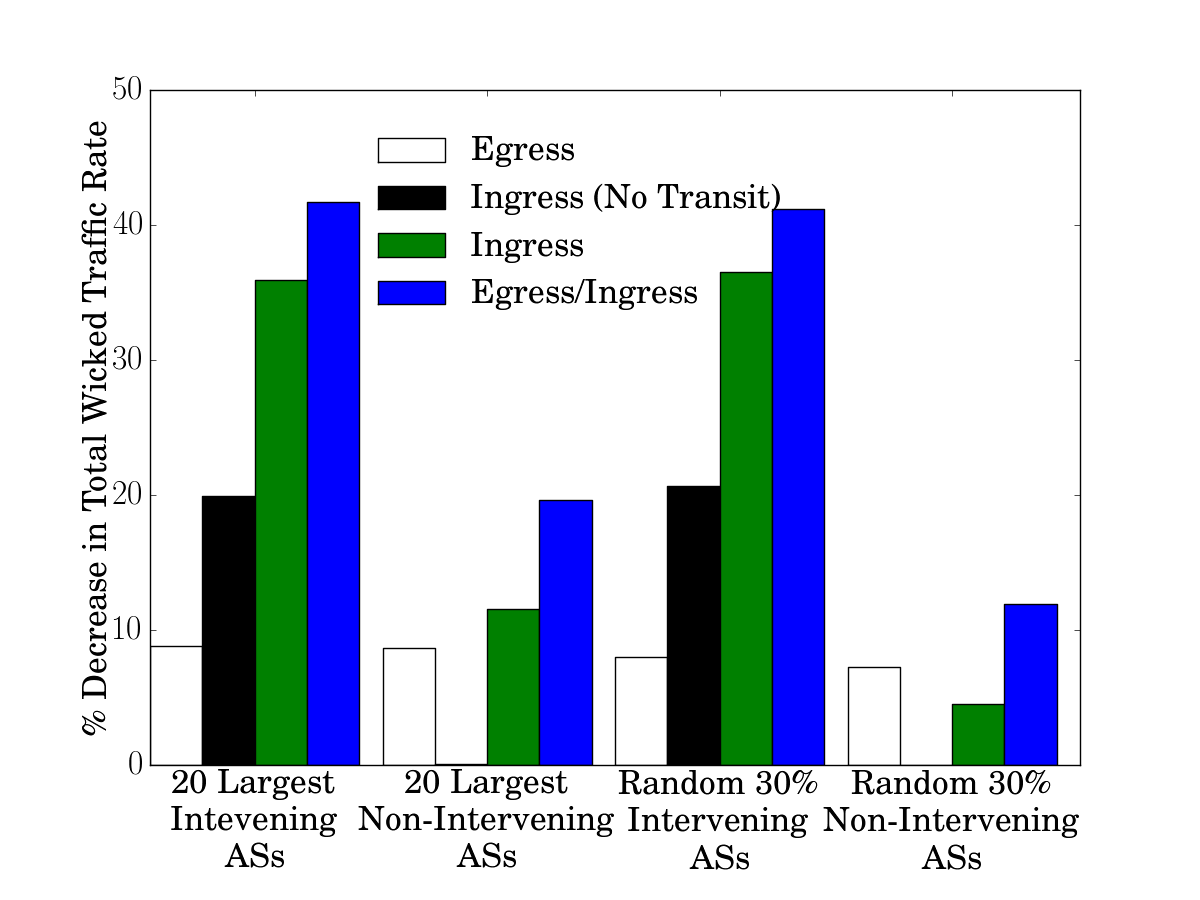}{The impact of interventions on wicked traffic rate on those ASes that
  intervene, and those that do not. ``20 largest'' is the effect when the 20
  largest ASes intervene; ``random $x$\%'' is the effect when $x$ percent of all
  ASes intervene.}

Most policy interventions under discussion have focused on ISPs' remediating
customer infections, which is akin to egress filtering. While this does reduce
wicked traffic levels, our results suggest that resources might be put to
better use by filtering incoming and transit traffic for wickedness.

Figure~\ref{fig:Fig4} compares the decrease
in wicked traffic at ASes that implement the interventions to the reduction at
ASes that do not adopt any interventions. The benefits for non-intervening ASes
represent a way to measure the positive externalities of security interventions
in the network. As expected, filtering egress traffic creates substantial
positive externalities, with non-intervening ASes experiencing similar
reductions in wicked traffic rates as intervening ASes. This effect holds for
both the largest ASes and a random selection of ASes. By contrast, filtering
ingress traffic has positive externalities only if wicked transit traffic is
blocked. In this case, the greatest benefits accrue to the intervening
ASes. This indicates that when filtering ingress traffic, the incentives for
adopting countermeasures are more aligned, and there should be less fear of
free-riding.

Furthermore, the positive externalities of ingress filtering (including transit
traffic) can vary greatly depending on which ASes intervene. The benefits to
non-intervening ASes are more than twice as large when the largest ASes
intervene rather than when ASes unilaterally intervene at random. This is
because large ASes attract more transit traffic, and so their filtering has a
greater impact.

\fig{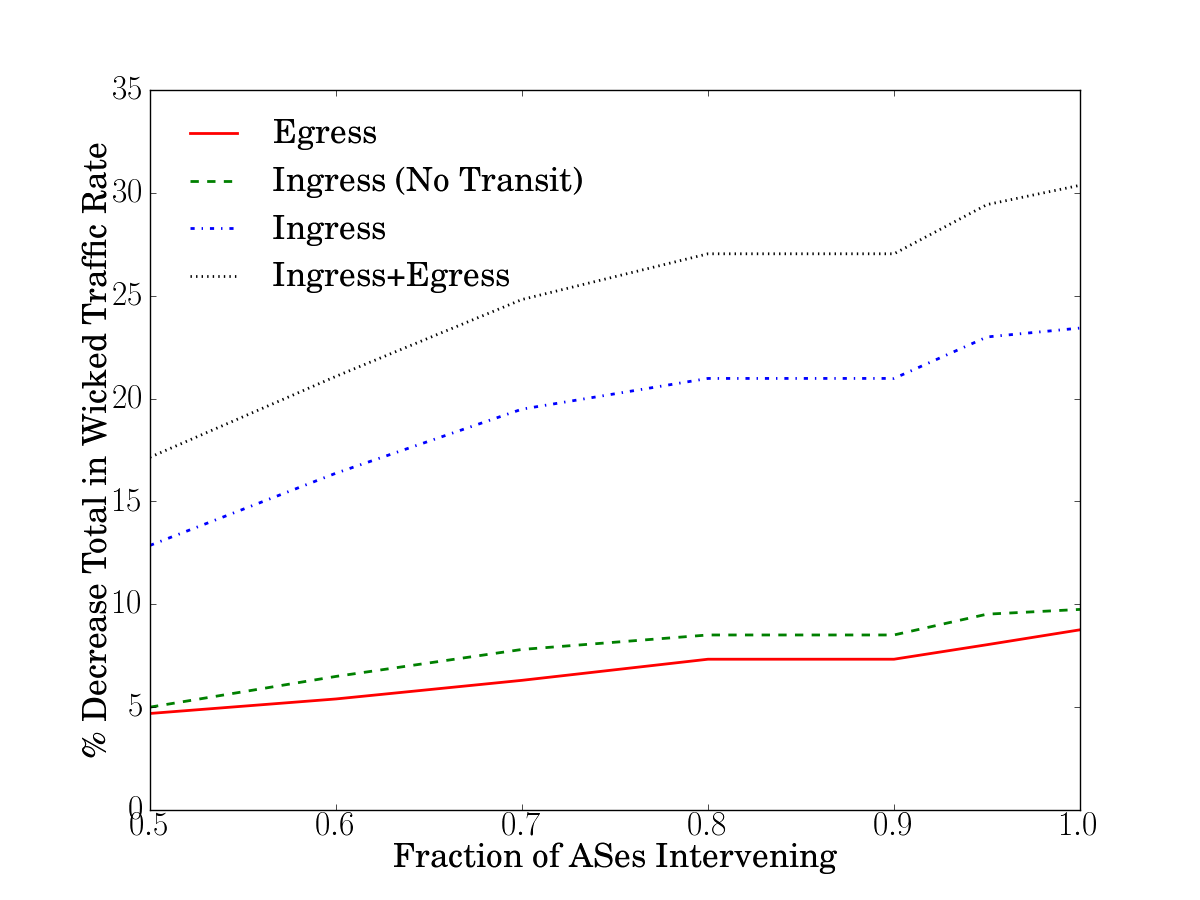}{The effect of the intervention of a fraction of the largest ASes. }

Even if having the largest ASes implement an intervention is the preferred
strategy for reducing wicked traffic on the Internet, it may not be possible to
enlist the support of all ASes. For example, even if all large US-based ISPs
adopted ingress and egress filtering, operators in other countries might choose
not to participate. To investigate the impact of incomplete adoption,
Figure~\ref{fig:Fig5} explores how varying the
proportion of large ASes that participate in the intervention affects the
reduction of malicious traffic.

Although wicked traffic falls as more ASes participate, the effect is
non-linear. For example, the differences between 80\% and 100\% of ASes
intervening are not great (from 27\% to 30\% wicked traffic reduction, an 11\%
change), whereas the differences between 60\% and 80\% are much greater (from
21\% to 27\%, a 29\% change). This suggests that country-level interventions are
much more likely to be effective if they include the majority of large ASes. For
example, if the all the largest ISPs based in the US were to intervene, that
would constitute at least 75\% of all large ASes.

\fig{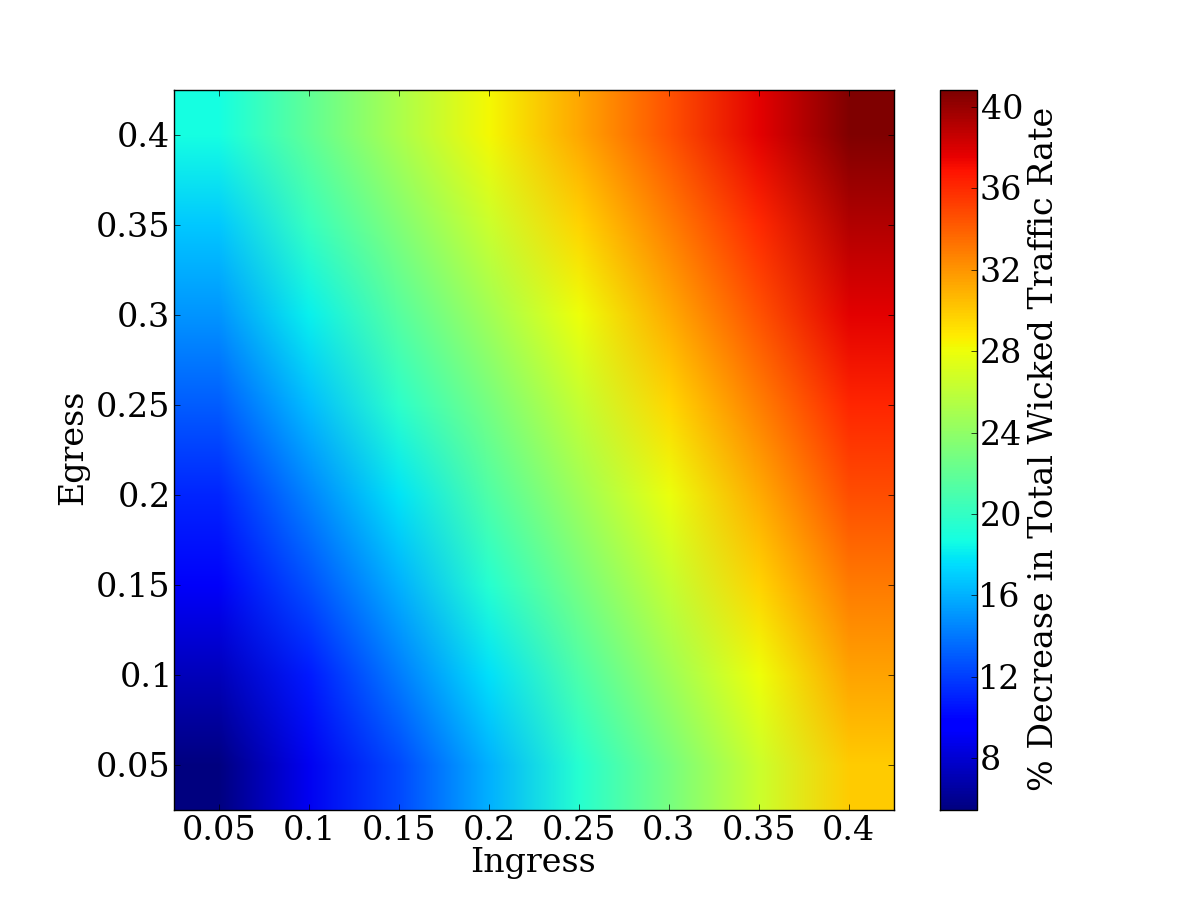}{The change in wicked traffic rate when varying the success rate of
  ingress and egress filtering. The scale indicates on the right the reduction
  in wicked traffic, from 0 to 40\%.}

In all the experiments reported previously, the ingress and egress filtering
effectiveness was set at 20\%.  However, some interventions are likely to be
more effective than others.  Notification-based schemes will filter less egress
wicked traffic than active quarantine, and increased data sharing could raise
the success rate of both ingress and egress filtering. It is very difficult to
get reliable information on the efficacy of these different approaches. Instead,
in Figure \ref{fig:Fig6} we explore how different
combinations of values for the success rates of ingress and egress filtering
affect the wicked traffic rates.  Ingress filtering is consistently more
effective at reducing overall wickedness.  For instance, ingress filtering 35\%
of wicked traffic and no egress traffic reduces the wicked traffic rate by the
same amount as 20\% ingress and 40\% egress filtering.

We also study the more aggressive intervention of completely blocking all
traffic originating from blacklisted ASes with unacceptably high wicked traffic
rates. Blacklisting results in a trade-off between reducing wicked traffic and
collateral damage caused by blocking innocent traffic. We consider only the case
where interventions are carried out by the 20 largest ASes (those of degree $\ge
170$), because, as seen previously, interventions are most successful when the
largest ASes act in concert.

\fig{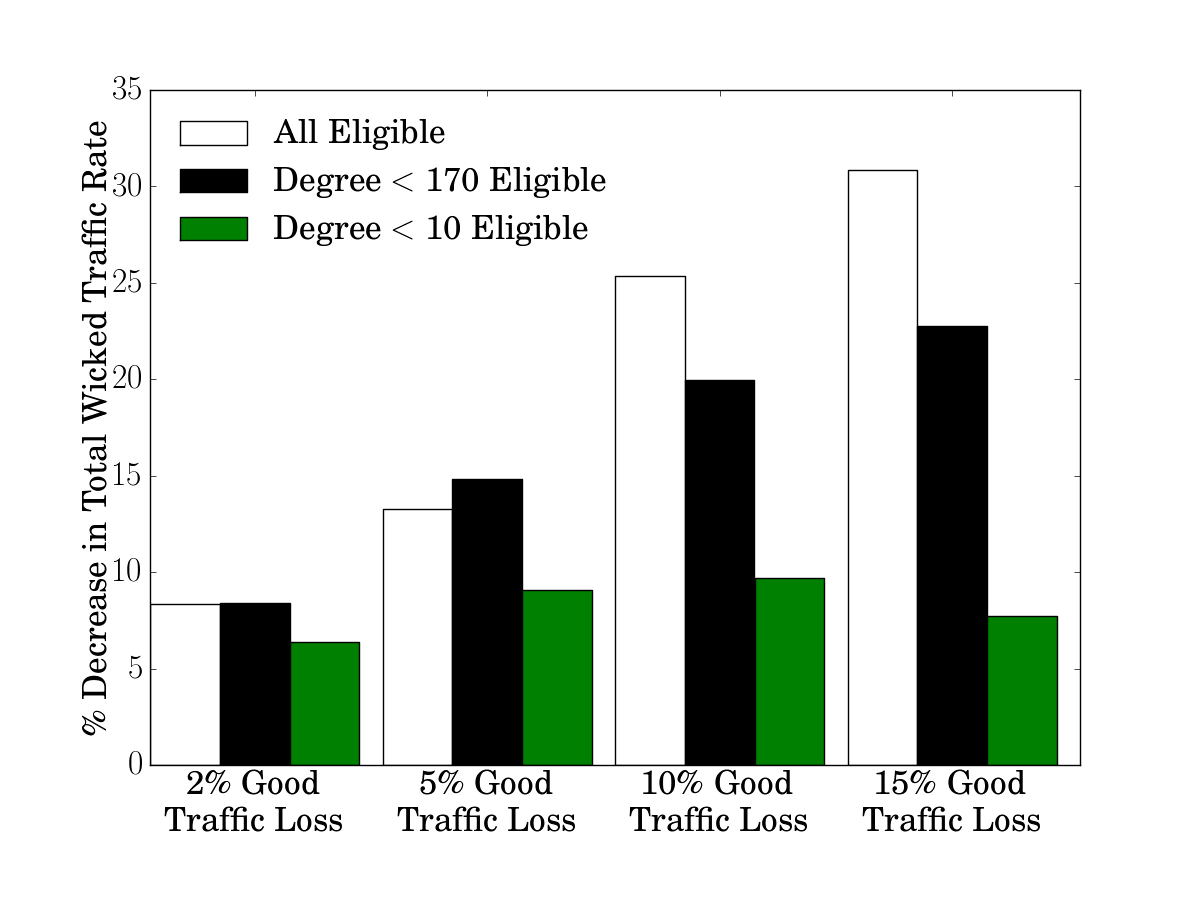}{The trade-off between reducing wicked traffic and losing legitimate
  traffic when blacklisting.}

There are two choices to make when applying blacklisting: first, the selection
of the level of wickedness above which ASes are blacklisted, and second, the
selection of whether to not blacklist larger ASes. We explore three levels of AS
size: blacklisting all ASes above the wickedness level, or those of degree
$<170$, or those of degree $<10$. For each choice of AS size, we select levels
of wickedness that result in losses of legitimate (good) traffic of 2\%, 5\%,
10\% and 15\%.

\fig{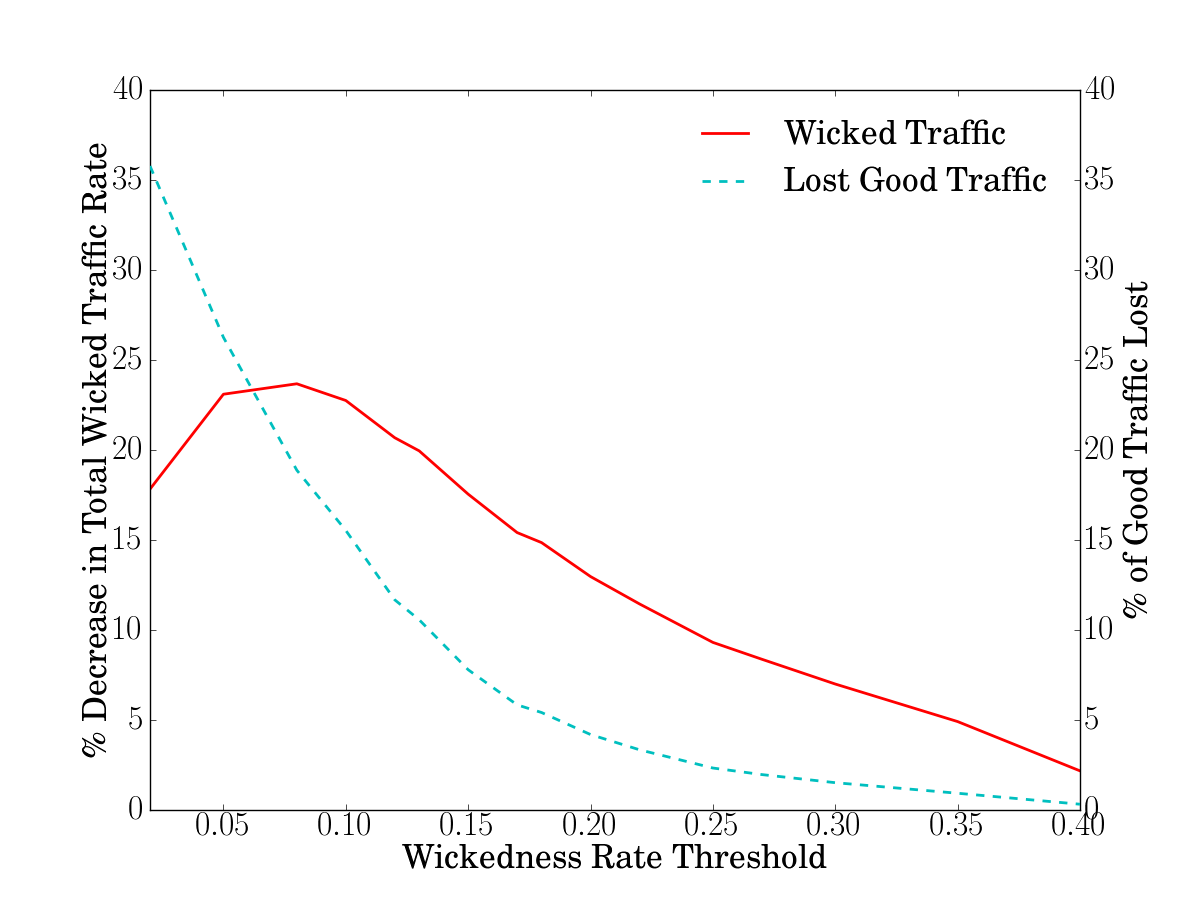}{The reduction in wicked traffic and the
 loss of legitimate (good) traffic when blacklisting all ASes of degree
 $<170$.} 

Figure~\ref{fig:Fig7} shows that the best strategy when
applying blacklisting depends very much on the level of legitimate traffic loss
we are willing to tolerate. For very low losses (2\%) the strategies have
similar results. For more moderate losses (5\%), we should blacklist all but the
20 largest ASes. Beyond that, it is more effective to blacklist all
ASes. However, we see diminishing returns as the level of acceptable loss
increases. For example, when blacklisting all ASes, a 50\% increase in
acceptable loss, from 10\% to 15\%, only reduces the wicked traffic by an
additional 23\%.

In fact, increasing the level of acceptable loss does not always reduce wicked
traffic. As can be seen in Figure~\ref{fig:Fig8},
the largest reduction of wicked traffic happens around a wickedness level of
$0.08$. Furthermore, there is a range over which the wicked traffic reduction
changes little; thus, the best choice of wickedness level would probably be
around $0.12$ for this example; anything lower increases the loss of legitimate
traffic with no beneficial wicked traffic reduction.

\subsection{Impact on Network Growth}
\label{sub_sec:ImpactOnNetworkGrowth}

The effect of malicious activity on the growth of the AS network is a complex
issue, one that we do not have the space to investigate in depth in this
paper. As an illustration of some of the potential for modeling chronic attacks
in ASIM, we briefly consider how the cost of intervention influences network
growth. Blacklisting is the simplest intervention to incorporate into the
economics of ASIM, because ASes earn money according to how much traffic they
route. Blacklisting reduces the amount of traffic (both legitimate and wicked)
seen by ASes and hence should change the evolution of the network. 

We carried out experiments where the 20 largest ASes intervene to blacklist all
traffic originating from ASes of degree less than 170. We set the wickedness
level for blacklisting to be $0.18$, which results in moderate legitimate
traffic loss. At this level, according to
Figure~\ref{fig:Fig7}, the best strategy is to blacklist
all sufficiently wicked ASes of degree less than 170.

Figure~\ref{fig:Fig9} shows how wicked traffic and lost
legitimate traffic change as the network evolves from 5\,000 to 13\,000 ASes. The
wicked traffic increases slightly (by about 9\%) and the lost legitimate traffic
decreases significantly (by about 66\%). To understand why this happens,
consider two classes of ASes: those that lose incoming traffic due to
blacklisting (class A) and those that do not (class B). In ASIM, every AS
depends on traffic for revenue, and so ASes in class A will earn less and hence
grow more slowly than ASes in class B. The ASes in class A will have reduced
levels of wicked traffic and increased levels of lost legitimate traffic
compared to those in class B. Thus, as ASes in class B grow more than those in
class A, the overall level of wicked traffic will increase, and the overall
level of legitimate traffic lost will decrease. This is exactly what we see in
Figure~\ref{fig:Fig9}. 

\fig{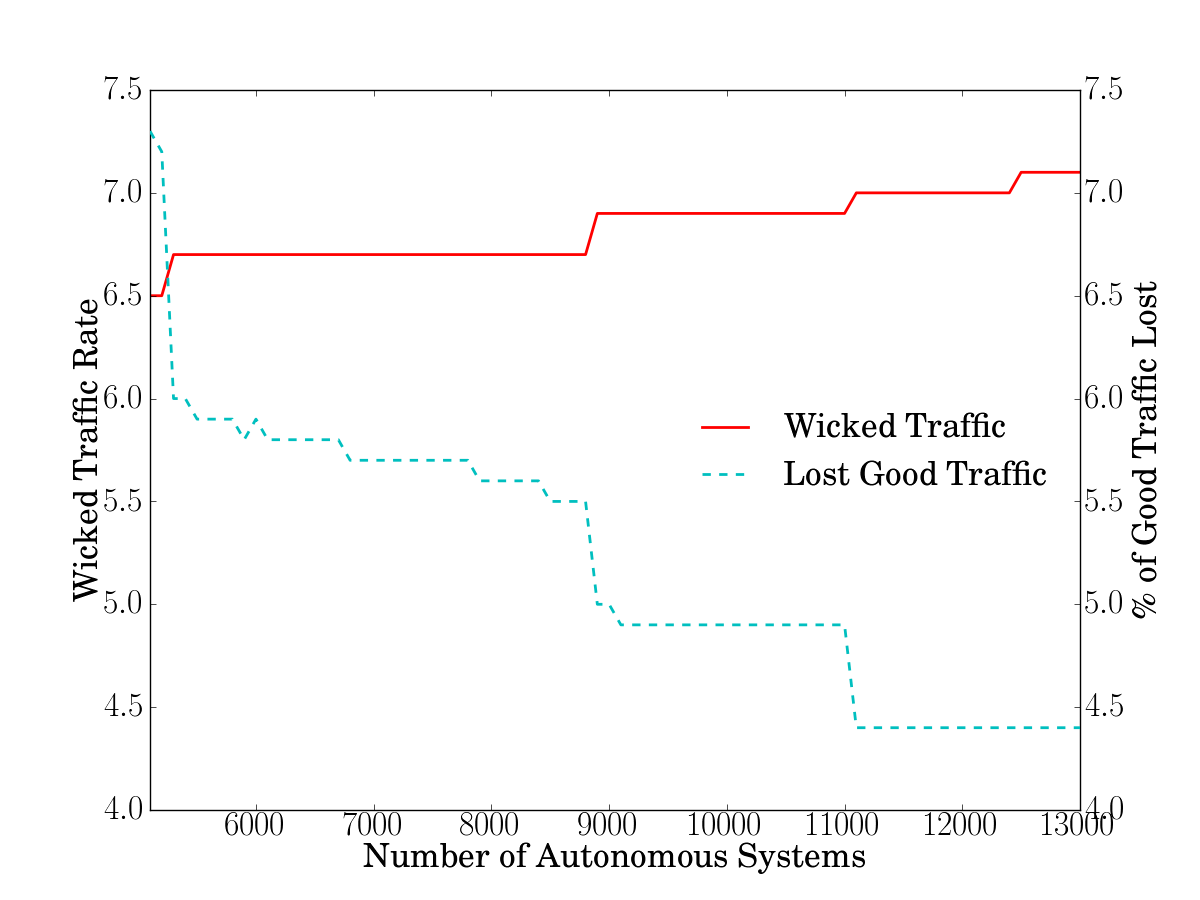}{The change in wicked traffic and loss of legitimate traffic over time
  as the network grows from 5\,000 to 13\,000 ASes. The wicked traffic rate is
  the percentage of all traffic that is wicked.}

Although blacklisting tends to promote ASes that receive more wicked traffic,
the rate at which wicked traffic increases is much slower than the rate at which
lost legitimate traffic decreases. Hence, blacklisting could still be considered
a viable strategy for reducing overall wickedness, at least in the short term.
Persuading individual ASes to voluntarily adopt blacklisting, however, would be
hard. Mandatory participation would likely be necessary.

\section{Related Work}
\label{sec:related}

Few studies have modeled the costs and benefits of
intervention to prevent the spread of malware across a
network. LeLarge~\cite{Lelarge2009,Lelarge09IEEE} used an agent-based
model to investigate the economics of interventions that counter
the spread of malware. However, LeLarge's model is much more abstract
than ASIM: agents exist on a random network, over which there is a
probabilistic spread of infections.  Agents can choose either to secure
themselves (at a cost) or to remain unsecured and risk loss. There is
no notion of geography or traffic. Varian~\cite{Varian04} proposed a
game-theoretic model to understand how security impacts the decisions
of other rational actors, but without considering network topology or
how infections may spread.  Subsequently, a number of
authors~\cite{Omic09,Aspnes07} have proposed models of
computer-infection spread that combine game theory with network
topology. These models focus on optimal strategies to combat a binary state of
infection. 

By contrast, a number of models have been developed to explore the spread of
malware, such as computer worms~\cite{Fei09}. Compartmental models of disease
spread (whether biological or electronic) are attractive methods for
investigating the progress of epidemics~\cite{AndersonMay92}. For example,
Ajelli et al. describe the spread of a botnet using such a
model~\cite{Ajelli2010}. Other work incorporates additional factors into
differential equation models, such as locations based on time
zone~\cite{Dagon06} and peer-to-peer protocols~\cite{Schafer08}. These
approaches focus on the spread of a single type of malware, such as a particular
worm or botnet. By contrast, our approach is to model all malware in a generic
way, incorporating both the economics of interventions, and the way
interventions affect the spread of malicious traffic on the Internet topology at
the AS level.

A major difference between agent-based models, such as ASIM, and differential
equation models, such as those described above, is that the latter assume that
populations are `well-mixed'; consequently they do not capture the effect of
skewed network topologies. Various extensions, such as percolation methods and
generating functions~\cite{Newman2002}, have been proposed as a method for
overcoming this limitation, spawning a great deal of interest in epidemics on
network topologies~\cite{Ganesh05}. Other extensions include using packet-level
data generated by computer network traffic simulators~\cite{Wei05}. In addition
to investigating the spread of malware across network topologies, mitigation
strategies such as quarantining malicious
hosts~\cite{Moore03,Palmieri2008,Coull05} have been investigated. However, to
the best of our knowledge, there are no studies that use these models to
investigate intervention policies at the ISP or Internet-level.

\section{Discussion}
\label{sec:discussion}

ASIM simplifies many aspects of routing on the real Internet. For example,
traffic in ASIM always follows the shortest path, whereas real traffic is also
influenced by agreements between ASes, following various conventions such as the
``valley free'' rule.  In ASIM ASes earn money from all traffic they route,
whereas in reality ASes earn money from their customers and pay their own
upstream providers. But we found in preliminary investigations that these added
complexities do not improve the accuracy of the model, at least in terms of
measures such as average path length, degree distribution, etc.  More detailed
modeling is a topic for future research and may lead to have implications for
the study of policy interventions.

Other model enhancements would allow us to study more carefully the impact of
interventions on the economics of network growth. We have presented a simple
initial approach, using blacklisting, but in future we intend to explore other
aspects, such as the cost of carrying out various interventions.  Blacklisting
is simple in that packets from a particular source are dropped, whereas
filtering only wicked traffic would likely be much more expensive, requiring a
sophisticated intrusion detection system (IDS).  Because of the performance
requirements, it may be infeasible to filter traffic using an IDS at the level
of the powerful routers used in the largest ASes. In this case, blacklisting and
improving end-user security may be the only reasonable options.

In our experiments with network growth, we kept the level of wickedness, or
compromised hosts, constant. This is clearly unrealistic as the number of
compromised hosts changes over time as some are cleaned up and others
infected. Furthermore, we expect that the amount of wicked traffic reaching
end-users will also influence infection rates. It is difficult to find good data
on how these rates change over time, and so it will be difficult to validate a
model that captures these aspects. One topic for future research is to model
dynamic wickedness levels, perhaps following an epidemiological model where
there is some rate of recovery from infection, and some rate of reinfection,
which is to some degree dependent on wicked traffic flow.

\section{Conclusions}
\label{sec:conclusion}

The results of our experiments using ASIM indicate that when filtering wicked
traffic, the best targets for intervention are a small group of the largest
ASes. Specifically, we find that intervention by the top 0.2\% of ASes (in terms
of size) is more effective than intervention by a randomly chosen subset of 30\%
of all ASes. However, we show that this efficacy rapidly drops off if less than
three quarters of that top 0.2\% intervene. This is an issue of importance if
not all the largest ASes fall within the same regulatory domain, such as a
nation-state.

Our experiments also illustrate the relative effectiveness of filtering ingress
and egress traffic. We show that filtering ingress traffic (including transit)
is more than twice as effective as filtering egress traffic
alone. Unsurprisingly, the effect of filtering is felt most strongly by those
actively filtering the data, although positive externalities can be seen if
outgoing or transit traffic is filtered. In our model, filtering egress traffic
is also a proxy for end-user remediation, which suggests that the current focus
on cleaning up ISP customers is not the most effective strategy.

In the case of blacklisting, we show that the choice of which ASes should be
exempt from blacklisting depends on how much legitimate traffic loss we are
willing to tolerate. If moderate levels of legitimate traffic loss are
acceptable, then large ASes should be exempt; however, if higher levels of
traffic loss are acceptable all ASes should be eligible for blacklisting. The
threshold for which ASes are blacklisted does not relate linearly to the
reduction in the wicked traffic rate. This is likely due to attrition of good
traffic, raising the fraction of wicked traffic seen. 

Our investigations of the impact of interventions on the evolution of the
network are brief and are limited to modeling the effect of blacklisting traffic
on growth. We show that blacklisting traffic results in a gradual increase in
wicked traffic, and a more rapid reduction in the loss of legitimate
traffic. Although this is beneficial in the short term, in the long-term those
ASes that profit most from wicked traffic will prosper at the expense of more
secure ASes, and so global effectiveness will decline.

We believe that the results reported in this paper are a good proof-of-concept
demonstration of how agent-based modeling can be useful to policy makers when
considering different interventions. We hope in future that our approach will
provide additional interesting results and tools to help policy makers
determine the best way to respond to the growing malware threat.

\begin{acknowledgement}
The authors gratefully acknowledge the support of DOE grant DE-AC02-05CH11231.
Stephanie Forrest acknowledges partial support of DARPA (P-1070-113237), NSF
(EF1038682,SHF0905236), and AFOSR (Fa9550-07-1-0532).
\end{acknowledgement}


\end{document}